\documentclass[
    ,final            
  ]
  {aipproc}

\layoutstyle{8x11double}

\def\lsim{\raise0.3ex\hbox{$\;<$\kern-0.75em\raise-1.1ex\hbox{$\sim\;$}}}
\def\gsim{\raise0.3ex\hbox{$\;>$\kern-0.75em\raise-1.1ex\hbox{$\sim\;$}}}
\def\barr{\begin{eqnarray}}
\def\earr{\end{eqnarray}}
\def\beq{\begin{equation}}
\def\eeq{\end{equation}}

\def\pbar{\overline{p}} 

\newcommand{\ba}{\begin{array}{c}}
\newcommand{\ea}{\end{array}}

\def\nue{{\nu_e}}
\def\nuebar{{\bar{\nu}_e}}

\def\nux{{\nu_x}}

\def\pvec{{\rm \bf p}}
\def\rvec{{\rm \bf r}}
\def\Pvec{{\rm \bf P}}
\def\Bvec{{\rm \bf B}}

\begin{document}

\title{Neutrinos from a core collapse supernova
\footnote{Plenary talk at NuFact07}
}

\classification{
14.60.Pq,       
97.60.Bw}       

\keywords{supernova, neutrino mixing, oscillations}

\author{Amol Dighe}{
  address={Tata Institute of Fundamental Research,
Homi Bhabha Road, Colaba, Mumbai 400005, India}
}

\begin{abstract}
The neutrino burst from a galactic supernova can help 
determine the neutrino mass hierarchy and $\theta_{13}$,
and provide crucial information about supernova astrophysics.
Here we review our current understanding of the neutrino
burst, flavor conversions of these neutrinos,
and model independent signatures of various 
neutrino mixing scenarios. 
\end{abstract}

\maketitle


Neutrino flavor 
conversions inside a SN are sensitive to extremely small 
$\theta_{13}$ values and the nature of neutrino mass hierarchy
\cite{dighe-smirnov}.
The observation of the neutrino burst from a galactic SN
will therefore provide information complementary to 
that from a long baseline experiment.
It will shed light on many of the outstanding questions
in neutrino oscillation physics and astrophysics.

In this article, we shall follow the production, propagation
and detection of SN neutrinos, which are akin to the 
source, the long baseline and the far detector of a 
neutrino factory setup.


\section{Operation of the SN $\nu$-factory}

\subsection{Core collapse and SN explosion}

Neutrinos and antineutrinos of all species are produced inside the
SN through pair production processes.
In addition, $\nue$ is also produced by electron capture on
protons: $p e^- \to n \nue$.
Before the collapse, neutrinos of all species are trapped
inside their respective ``neutrinospheres''
around $\rho \sim 10^{10}$g/cc.

When the iron core reaches a mass close to its Chandrasekhar limit,
it becomes gravitationally unstable and collapses.
A hydrodynamic shock is formed when the matter reaches nuclear
density and becomes incompressible.
When the shock wave passes through the
$\nue$ neutrinosphere, a short $\nue$
``neutronization'' burst is emitted, which
lasts for $\sim$10 ms.
The object below the shock wave, the ``protoneutron star,'' then
cools down with the emission of neutrinos of all species.
This emission takes place over a time period of $t \sim 10$ s
\cite{book}.

The eventual explosion of the star involves the stalling of the
original shock wave, its revival by the trapped neutrinos, and
a ``delayed'' explosion where large scale convections play
an important role \cite{bethe,janka}.
However, for the $~\sim 10$ sec neutrino burst that we focus
on here, the actual explosion mechanism, or even whether
the star successfully explodes or not, is mostly immaterial.

\subsection{The source: primary neutrino fluxes}
\label{source}

A SN core acts essentially like a neutrino blackbody source
with flavor-dependent fluxes.
Since the fluxes are almost identical for
$\nu_\mu, \nu_\tau, \bar{\nu}_\mu$ and $\bar{\nu}_\tau$, 
all these species may be represented by $\nux$.
The ``primary fluxes'' $F^0_{\nu_\alpha}$
may be parametrized by the total number fluxes 
$\Phi_0(\nu_\alpha)$, average energies 
$\langle E_0(\nu_\alpha) \rangle$, and the ``pinching
parameters'' that characterize their spectral shapes
~\cite{keil1}.

The values of the parameters are highly model dependent,
as can be seen from Table~\ref{models}.
The two leading models -- the Livermore simulation 
\cite{livermore} and the more recent Garching 
calculation \cite{garching} -- agree on
$ {\langle E_0(\nue) \rangle} \approx 12$ MeV and
${\langle E_0(\nuebar) \rangle} \approx 15$ MeV,
and have consistent values for the pinching parameters,
but they differ widely on ${\langle E_0(\nux) \rangle}$ 
and the ratios of total fluxes.
In particular, the equipartition of energy
assumed in the Livermore model
is not a feature of the Garching model.

\begin{table}[h]
\begin{tabular}{lccc}
\hline
{Model} & 
${\langle E_0(\nux) \rangle}$ &
$\frac{\Phi_0(\nue)}{\Phi_0(\nux)}$ &
$\frac{\Phi_0(\nuebar)}{\Phi_0(\nux)}$\\
\hline
{Garching} &
{18} & {0.8 }& {0.8} \\
{Livermore} & 
{24 }& {2.0}&{1.6} \\
\hline
\end{tabular}
\caption{Differences in flux predictions from SN models}
\label{models}
\end{table}

In the light of the model dependence, it is important to make sure
that the inferences drawn from the observed neutrino spectra
do not depend strongly on the exact model parameters.

\section{Flavor transformations along the long
baseline}

Neutrinos that are produced approximately as
mass eigenstates at very high ambient matter density in the core
propagate outwards from the neutrinosphere.
They have to travel
through the core, mantle and envelope of the star,
through the interstellar space, and possibly even through
some part of the Earth before arriving at the detector.
Inside the SN, collective and MSW matter effects take place.
In the interstellar space, neutrino mass eigenstates travel
independently, whereas oscillations, enhanced by MSW effects,
occur inside the Earth.

\subsection{MSW resonances inside the SN}

The traditional picture of flavor conversions in a SN is based on
the assumption that the effect of neutrino-neutrino interactions
is small.
In this case, flavor conversions occur most efficiently
in the MSW resonance regions.
SN neutrinos must pass through
two resonance layers: the H-resonance layer at
$\rho_{\rm H}\sim 10^3$ g/cc characterized by
$(\Delta m^2_{\rm atm}, \theta_{13})$,
and the L-resonance layer at
$\rho_{\rm L}\sim 10$ g/cc characterized by 
$(\Delta m^2_{\odot}, \theta_{12})$.
This hierarchy of the resonance densities, along with their
relatively small widths, allows the transitions in the two
resonance layers to be considered independently
\cite{kuo-rmp}.

The outcoming incoherent mixture of vacuum mass eigenstates is
observed at a detector as a combination of primary fluxes
of the three neutrino flavors:
\begin{eqnarray}
F_{\nue} & = & p F_{\nue}^0 + (1-p) F_{\nux}^0 ~, \\
F_{\nuebar} & =  & \pbar F_{\nuebar}^0 + (1-\pbar) 
F_{\nu_x}^0 ~,
\label{feDbar}
\end{eqnarray}
where $p$ and $\pbar$ are the
survival probabilities of $\nue$  and $\nuebar$ respectively.
 
\begin{table}[h]
\caption{Survival probabilities for neutrinos, $p$, and
antineutrinos, $\pbar$, in various mixing scenarios
\label{tab-pbar}}
\begin{tabular}{llccc}
\hline
 &
Hierarchy &  $\sin^2 \theta_{13}$  &  $p$ &  $\pbar$ \\
\hline
A & Normal & $\gsim 10^{-3}$  & 0  & $\cos^2\theta_\odot$ \\
B & Inverted &  $\gsim 10^{-3}$ &  $\sin^2\theta_\odot$ &  0 \\
C & Any & $\lsim 10^{-5}$  & $\sin^2\theta_\odot$
&  $\cos^2\theta_\odot$ \\
\hline
\end{tabular}
\end{table}

The neutrino survival probabilities are governed by
the adiabaticities of the resonances traversed, 
which are directly
connected to the neutrino mixing scheme.
In particular, whereas the L-resonance is always adiabatic and
appears only in the neutrino channel, the adiabaticity of
the H-resonance depends on the value of $\theta_{13}$, and
the resonance shows up in the neutrino (antineutrino) channel
for a normal (inverted) mass hierarchy.
Table~\ref{tab-pbar} shows the survival probabilities
in various mixing scenarios.
For intermediate values of $\theta_{13}$, i.e.
$10^{-5}\lsim\sin^2 \theta_{13} \lsim 10^{-3}$,
the survival probabilities depend on energy as well
as the details of the SN density profile
\cite{dighe-smirnov}.

Scenarios A, B and C are the ones that can in principle be
distinguished through the observation of a SN neutrino burst.
Note that the sensitivity to $\theta_{13}$ is comparable to
the expected reach of a neutrino factory.

\subsection{Collective effects at large $\nu$ densities}
\label{collective}

The neutrino and antineutrino densities near the neutrinosphere 
are extremely high ($10^{30-35}$ per cm$^3$), which
make the $\nu-\nu$ interactions in this region significant
\cite{duan-fuller-carlson-qian-0606616,duan-fuller-carlson-qian-0608050}.
Indeed, the Hamiltonian is now given by
\beq
H(\pvec, \rvec) = H_{vac}(p)+V(\rvec)+H_{\nu\nu}(\pvec,\rvec) \; ,
\eeq
where $H_{vac}$ is the vacuum Hamiltonian, $V(\rvec)$ is the
MSW potential due to electrons and the $\nu-\nu$ interaction
potential $H_{\nu\nu}$ is 
\cite{thompson-mckellar-PLB259,raffelt-sigl-NPB406,thompson-mckellar-PRD49}
\beq
H_{\nu \nu}(\pvec,\rvec) = 
\sqrt{2}G_F\int \frac{d^3{\bf q}}{\left(2\pi\right)^3}
\kappa_{\pvec {\bf q}} 
(n_{\nu}\rho -
\bar{n}_{\nu} \bar{\rho})~.
\label{Hnunu}
\eeq
Here $n_\nu({\bf q},\rvec,t)$ and $\rho({\bf q},\rvec,t)$
are the number density and density matrix of neutrinos with
momentum ${\bf q}$, whereas 
$\kappa_{\pvec {\bf q}} \equiv 
(1-\cos \theta_{\pvec {\bf q}})$. 
The quantities with a ``bar'' represent
antineutrinos. 
Note that the evolutions for $\nu$ and $\bar{\nu}$
are nonlinear, and are coupled to each other.

The distinctive features in the flavor evolution of such 
a relativistic gas have been identified in
\cite{samuel-PRD48,kostecky-samuel-9506262,pantaleone-PRD58,samuel-9604341}.
The evolutions of the density matrices of neutrinos and
antineutrinos may be represented in terms of the precessions
of the corresponding Bloch vectors, termed as ``polarization
vectors'' $\Pvec$.
The traditional MSW oscillations correspond to $\nu$ and 
$\bar\nu$ of each energy independently precessing about
$\Bvec$, the Bloch vector corresponding to the Hamiltonian
$H_{vac}(p) + V(\rvec)$.

Different collective effects play important roles in different
regions of the star \cite{fogli-lisi-marrone-mirizzi-0707.1998}.
When the neutrino density is extremely high,
$\Pvec$'s of all energies remain tightly bound
together and precess with a common frequency,
giving rise to synchronized oscillations
\cite{pastor-raffelt-semikoz-0109035}.
At lower densities, the $\Pvec$'s remain bound together to a 
large extent, but have a tendency to relax to the state that 
has the lowest energy, causing bipolar oscillations 
\cite{hannestad-raffelt-sigl-wong-0608095,duan-fuller-carlson-qian-0703776}.
Collective interactions also predict ``spectral split,''
a complete swapping of the energy spectra of two
neutrino flavors above or below a critical energy,
as the neutrinos transit from a
region where collective effects dominate to a region where
the neutrino density is low
\cite{raffelt-smirnov-0705.1830,raffelt-smirnov-0709.4641}.

The analytic treatment of collective effects till now has been
mostly in the two-flavor limit, assuming a steady-state, 
spherical, half-isotropic, finite source.
The dependence of the flavor evolution on the direction of
propagation of the neutrino may give rise to direction-dependent
evolution \cite{duan-fuller-carlson-qian-0606616,duan-fuller-carlson-qian-0608050},
or to decoherence effects
\cite{pantaleone-PRD58,raffelt-sigl-0701182}, but
for a realistic asymmetry between neutrino and antineutrino
fluxes, such effects are likely to be small
\cite{fogli-lisi-marrone-mirizzi-0707.1998,estebanpretel-pastor-tomas-raffelt-sigl-0706.2498}
and a so-called ``single-angle'' approximation  can be used.
Recently, a formalism for analyzing the three-flavor effects
has also been developed \cite{three-flavor}.

For SN density profiles where the collective effects are
over before the MSW effects begin, the collective effects
are equivalent to changing the primary spectra available for
further propagation, so that results in 
Table~\ref{tab-pbar} stay valid. 
For ``shallow'' density profiles \cite{o-ne-mg} 
where the collective and MSW regions may overlap, 
the situation is more complex
and has to be analyzed separately.

\subsection{Oscillations inside the Earth matter}
\label{earth}

If the neutrinos travel through the mantle, and possibly
core, of the Earth before reaching
the detector, the neutrinos undergo oscillations inside
the Earth and the survival probabilities change
\cite{cairo,ls2,sato}.
This change however occurs only in those scenarios
in Table~\ref{tab-pbar} where the value of the survival
probability is nonzero. 
This provides the means for discriminating between
various mixing scenarios.

When antineutrinos, for example, travel through the
mantle and the core of the earth,
the sharp density jumps give rise to the survival probability
of the form \cite{corewiggles}
\beq
\pbar^D \approx \cos^2 \theta_{12} + \sum_{i=1}^7 \bar{A}_i
\sin^2(\phi_i/2)
\label{pdcore}
\eeq
in the two-layer model of the Earth,
where the coefficients $ \bar{A}_i$ are functions of the mixing
angle $\theta_{12}$ in vacuum, mantle and core.
The phases $\phi_i$ depend on the distance $L$ traveled through
the Earth matter and the values of mass squared differences
$\overline{\Delta m^2}$ 
between $\bar{\nu}_1$ and $\bar{\nu}_2$ 
in the mantle and the core.
If the neutrinos traverse only through the mantle,
only one oscillating term is present in (\ref{pdcore}),
with 
$\phi = 2 \overline{\Delta m^2}_{\rm mantle} L/ E$ 
\cite{fourier}.

\section{Distinguishing between neutrino mixing scenarios}
\label{identify}

The only SN observed in neutrinos till now, SN1987A,
yielded only $\sim$20 events. Though it confirmed our
understanding of the SN cooling mechanism, the number
of events was too small to say anything concrete
about neutrino mixing
(see \cite{ls-sn87} and references therein).
On the other hand, 
If a SN explodes
in our galaxy at 10 kpc from the Earth,
we expect $\sim 10^4$ events at Super-Kamiokande (SK) 
through the inverse beta decay process $\nuebar p \to n e^+$.
This process, dominant at any water
Cherenkov or scintillation detector, will be
instrumental in determining the $\nuebar$ spectrum.
In order to measure the $\nue$ spectrum cleanly, however, 
one needs a large liquid Ar detector, 
with the relevant process
$\nu_e + ~^{40}{\rm Ar} \to ~^{40}{\rm K}^* + e^-$.

The uncertainties in the primary neutrino spectra 
make the task of determining the survival probabilities
$p$ and $\pbar$ almost impossible,
and alternative model independent signatures of
various neutrino mixing scenarios need to be looked for.
The Earth effects and shock wave
imprints seem particularly promising for this 
purpose.
Since the characteristics of the neutronization burst
are robust across models, the structure of the neutronization 
peak can also identify scenario A, where the peak is 
highly suppressed \cite{ricard-neutro}.

Earth effects manifest themselves in two ways.
Firstly, the total number of events and the spectral shape
changes.
Secondly, Earth effect oscillations are introduced,
which may be identified even at a single detector.

\subsection{Comparing signals at multiple detectors}
\label{multiple}
 
If neutrinos travel different distances through the
Earth before reaching two detectors, the difference in the
signals at the detectors could show evidence for Earth effects.
In order to obtain a statistically significant difference
in the neutrino spectra, the detectors need to be of the
size of SK or larger.

The IceCube detector, though designed to detect 
individual neutrinos with $E \gsim 150$ GeV, is able to
detect a SN neutrino burst during which
the number of Cherenkov
photons detected by the optical modules would increase
much beyond the background fluctuations \cite{halzen,halzen2}.
Though this does not measure energies of individual neutrinos,
the total luminosity can be determined at the per cent level.
Comparison of the luminosities as functions of time at the 
IceCube and SK (or its larger version) 
can identify the earth effects,
since they are typically time dependent \cite{icecube}.

The relative locations of SK and IceCube imply that
for the SN in a large portion of the sky,
neutrinos pass through the Earth for only one of the
detectors.
This makes the SK--IceCube comparison
an interesting prospect.

\subsection{Identifying Earth effects at a single detector}
\label{single}
 
The oscillating terms $\sin^2 (\phi_i/2)$ in (\ref{pdcore})
can manifest themselves as peaks in the Fourier power spectrum 
of the ``inverse energy'' spectrum of $\nuebar$
\cite{fourier}:
\beq
G_N(k) =  \left| \sum_{\rm events} e^{i k y_{\rm event}} \right|^2 / 
N_{\rm events} \; ,
\eeq
where $y \equiv (25 ~{\rm MeV})/E$.
The positions of these peaks are independent of the primary
neutrino spectra, being determined by
the solar oscillation parameters, 
the Earth matter density, and the position of the SN in the sky.
Therefore, Earth effects can be identified merely by
identifying the presence of these oscillation frequencies in the
observed spectrum.
If the neutrinos pass only through the mantle, there is only 
one Fourier peak.
When the core is also involved, three out of the seven
possible peaks may have significant power \cite{corewiggles}, 
leading to an easier identification of the Earth matter
effects.

Finite energy resolutions of detectors tend to
smear out the modulations in the energy spectrum,
and suppress high-$k$ peaks. 
The comparison between a simulated megaton water Cherenkov
detector and a 32 kt scintillation detector \cite{corewiggles}
shows that
the better resolution of the scintillator detector almost
compensates for the much larger water Cherenkov detector size.

The observation of Earth effects in $\nue$ $(\nuebar)$, 
either through the luminosity
comparison or through Fourier peaks,
eliminates scenario A (B) independently 
of SN models.

 \section{Neutrinos for SN astrophysics}
\label{astro}
 
\subsection{Pointing to the SN in advance}
\label{pointing}

Since neutrinos are expected to arrive hours before the optical
signal from the SN, the neutrino burst serves as an early warning
\cite{snews} to the astronomy community.
Being able to determine the position of the SN in the sky is
also crucial for determining the Earth crossing
path for the neutrinos in the absence of the SN observation
in the electromagnetic spectrum.

A SN may be located through the directionality of
the $\nu e^- \to \nu e^-$ elastic scattering events
in a water Cherenkov detector such as SK~\cite{beacom,ando}.
The directionality of this reaction is primarily limited
by the angular resolution of the detector,
the kinematical deviation of the final-state electron 
direction from the initial neutrino,
and the nearly isotropic $\nuebar p \to n e^+$
background which is 30-40 times larger than the signal. 
Adding to the water a small amount of gadolinium, 
an efficient neutron absorber, would
allow one to detect the neutrons and thus to tag 
the inverse beta reactions~\cite{gadzooks}.
Efficient neutron tagging can improve the pointing 
accuracy at SK from $\sim8^\circ$ to $\sim 3^\circ$ for a SN
at 10 kpc \cite{pointing}.
       
\subsection{Tracking the shock wave in neutrinos}
\label{shock}
  
The passage of the shock wave through the 
H-resonance ($\rho \sim 10^3$ g/cc) a
few seconds after the core bounce may
break adiabaticity, thereby modifying the spectral
features of the observable neutrino flux
~\cite{fuller,takahashi,ls1,lisi-shock}.
One expects a ``dip'' in $\langle E_e \rangle$ as well
as the number of events, and a simultaneous peak
in $\langle E_e^2 \rangle / \langle E_e \rangle^2$.
If a reverse shock is also present,
the above features become a a double-dip and a double-peak 
respectively \cite{revshock}.

Since the density of the H-resonance layer depends on energy, 
the positions of the dips in the number of events at different 
neutrino energies would allow one to trace the shock
propagation while it is in the mantle around densities of
$\rho \sim 10^3$ g/cc \cite{revshock}.

The shock wave effects can be diluted by stochastic density
fluctuations as well as turbulence.
For example, for $\theta_{13} \gsim 10^{-4}$ the shock
wave imprints may be partly erased with $\delta$-correlated 
stochastic fluctuations \cite{stochastic}.
If the turbulent convection generated behind the
shock wave is sufficiently large, flavor depolarization 
takes place at the H resonance, so that the
sharp shock wave effects are replaced by gradual 
depolarization effects \cite{friedland}.
A recent hydrodynamic simulation
\cite{brockman}
suggests that some of the shock wave effects survive
in spite of the smearing factors above.
Sterile neutrinos may leave their imprints in the shock wave
\cite{choubey-ross1}, which can also survive in the presence
of turbulence \cite{choubey-ross2}.

The nonmonotonic density profile of the shock wave may
cause the neutrinos to pass through multiple H resonances.
This gives rise to oscillations in the survival probabilities
of neutrinos, where the positions of maxima and minima are
independent of primary  fluxes \cite{phase}.
The oscillations are however smeared out due to the finite
energy resolutions of detectors, and the signal may
be detectable only in extremely optimistic cases.

The observation of any of the shock wave imprints
in the $\nue$ ($\bar\nu_e$) spectra would imply that the neutrino 
mixing scenario is A (B).

\section{Synergy between SN neutrinos and a neutrino factory}
\label{concl}

A galactic SN burst is a rare phenomenon, expected to occur
only once in a few decades.
However, its observation 
is expected to reap a rich scientific harvest.
It is therefore imperative that we are ready with suitable
long term detectors that will observe the relevant signals.
In the meanwhile, better theoretical understanding of
neutrino transport inside the SN, combined with more accurate 
measurements of the neutrino mixing parameters, will equip us for
making the most of the cosmic catastrophe.

Earth effects and shock wave imprints are robust signals
for specific neutrino mixing scenarios that are
unlikely to be mimicked by anything else.
The information available from them is in the form of
a combination of mass hierarchy and a $\theta_{13}$ range, 
so that complementary information from
long baseline experiments is also required.
If $\theta_{13}$ and mass hierarchy is already determined 
at terrestrial experiments, concrete information on the primary neutrino fluxes will be
obtained.
On the other hand, if the burst is observed before the
mixing parameters are measured, we shall have some advance
idea about the neutrino mixing parameters expected, and
that will guide our efforts towards the neutrino factory.

\begin{theacknowledgments}
We would like to thank the organisers of NuFact07 for
their hospitality. This work is partly supported through the
Partner Group program between the Max Planck Institute
for Physics and Tata Institute of Fundamental Research.

\end{theacknowledgments}

\end{document}